\documentclass[11pt,a4paper]{article}
\usepackage[noblocks]{authblk}
\usepackage[hyperref]{acl2021}
\usepackage{times}
\usepackage{latexsym}

\usepackage{microtype}

\usepackage{graphicx}
\usepackage{xspace}
\usepackage{xcolor}
\usepackage{url}
\usepackage{changepage}
\usepackage{bold-extra}
\usepackage{amsmath}
\usepackage{amssymb}
\usepackage{xspace}
\usepackage{soul}
\usepackage{float}

\aclfinalcopy
\def\aclpaperid{2165}

\newcommand\bioasq{\textsc{bioasq}\xspace}
\newcommand\IR{\textsc{ir}\xspace}
\newcommand\ir{\textsc{ir}\xspace}
\newcommand\qa{\textsc{qa}\xspace}
\newcommand\mrc{\textsc{mrc}\xspace}
\newcommand\MAP{\textsc{map}\xspace}
\newcommand\MRR{\textsc{mrr}\xspace}

\newcommand\idf{\textsc{idf}\xspace}
\newcommand\bmtf{\textsc{bm25}\xspace}
\newcommand\bmtfbmtf{\textsc{bm25+bm25}\xspace}
\newcommand\bilstm{\textsc{bilstm}\xspace}
\newcommand\mlp{\textsc{mlp}\xspace}
\newcommand\cnn{\textsc{cnn}\xspace}
\newcommand\rnn{\textsc{rnn}\xspace}

\newcommand\wordtovec{\textsc{word2vec}\xspace}
\newcommand\elmo{\textsc{elmo}\xspace}
\newcommand\bert{\textsc{bert}\xspace}
\newcommand\bertPipe{\textsc{bert+bert}\xspace}
\newcommand\jbert{\textsc{jbert}\xspace}
\newcommand\jbertnf{\textsc{jbert-nf}\xspace}
\newcommand\jbertadapt{\textsc{jbert-adapt}\xspace}
\newcommand\jbertadaptnf{\textsc{jbert-adapt-nf}\xspace}
\newcommand\bertbcnn{\textsc{bert+bcnn}\xspace}
\newcommand\bertpdrmm{\textsc{bert+pdrmm}\xspace}
\newcommand\pacrr{\textsc{pacrr}\xspace}
\newcommand\drmm{\textsc{drmm}\xspace}
\newcommand\pdrmm{\textsc{pdrmm}\xspace}
\newcommand\jpdrmm{\textsc{jpdrmm}\xspace}
\newcommand\bertjpdrmmnf{\textsc{bjpdrmm-nf}\xspace}
\newcommand\bertjpdrmmadapt{\textsc{bjpdrmm-adapt}\xspace}
\newcommand\bertjpdrmmadaptnf{\textsc{bjpdrmm-adapt-nf}\xspace}
\newcommand\bertjpdrmm{\textsc{bjpdrmm}\xspace}
\newcommand\pdrmmbcnn{\textsc{pdrmm+bcnn}\xspace}
\newcommand\pdrmmPipe{\textsc{pdrmm+pdrmm}\xspace}
\newcommand\bcnn{\textsc{bcnn}\xspace}

\newcommand\positdrmm{\textsc{posit-drmm}\xspace}
\newcommand\knrm{\textsc{knrm}\xspace}
\newcommand\convknrm{\textsc{convknrm}\xspace}
\newcommand\squad{\textsc{squad}\xspace}

\title{A Neural Model for Joint Document and Snippet Ranking in Question Answering for Large Document Collections}

\author[1,2]{Dimitris Pappas}
\author[1]{Ion Androutsopoulos}
\affil[1]{Department of Informatics, Athens University of Economics and Business, Greece}
\affil[1]{\url{pappasd@aueb.gr, ion@aueb.gr}}
\affil[2]{Institute for Language and Speech Processing, Research Center ‘Athena’, Greece}
\affil[2]{\url{dpappas@athenarc.gr}}

\date{}

\begin{document}
\maketitle
\begin{abstract}Question answering (\qa) systems for large document collections typically use pipelines that (i) retrieve possibly relevant documents, (ii) re-rank them, (iii) rank paragraphs or other snippets of the top-ranked documents, and (iv) select spans of the top-ranked snippets as exact answers. Pipelines are conceptually simple, but errors propagate from one component to the next, without later components being able to revise earlier decisions.
We present an architecture for joint document and snippet ranking, the two middle stages, which leverages the intuition that relevant documents have good snippets and good snippets come from relevant documents. The architecture is general and can be used with any neural text relevance ranker. We experiment with two main instantiations of the architecture, based on \positdrmm (\pdrmm) and a \bert-based ranker.

Experiments on biomedical data from \bioasq  show that our joint models vastly outperform the pipelines in snippet retrieval, the main goal for \qa, with fewer trainable parameters, also remaining competitive in document retrieval.
Furthermore, our joint \pdrmm-based model is competitive with \bert-based models, despite using  orders of magnitude fewer parameters. These claims are also supported by human evaluation on two test batches of \bioasq. 
To test our key findings on another dataset, we modified the Natural Questions dataset so that it can also be used for document and snippet retrieval. Our joint \pdrmm-based model again outperforms the corresponding pipeline in snippet retrieval on the modified Natural Questions dataset, even though it performs worse than the pipeline in document retrieval.
We make our code and the modified Natural Questions dataset publicly available. 
\end{abstract}

\section{Introduction} \label{sec:Introduction}

Question answering (\qa) systems that search large document collections \cite{Voorhees2001,Tsatsaronis_et_al_bioasq,Chen2017} typically use pipelines operating at gradually finer text granularities. A fully-fledged pipeline includes components that (i) retrieve possibly relevant documents typically using conventional information retrieval (\ir); (ii) re-rank the retrieved documents employing a computationally more expensive document ranker; (iii) rank the passages, sentences, or other `snippets' of the top-ranked documents; and (iv) select spans of the top-ranked snippets as `exact' answers.
Recently, stages (ii)--(iv) are often pipelined neural models, trained individually \cite{Hui_et_al_pacrr,Pang_et_al_reviewer_6,lee-etal-2018-ranking,McDonald_et_al_pdrmm,Pandey-et-al-infret-2019,mackenzie-etal-2020,sekuli2020longformer}. Although pipelines are conceptually simple, errors propagate from one component to the next \cite{hosein2019measuring}, without later components being able to revise  earlier decisions. For example, once a document has been assigned a low relevance score, finding a particularly relevant snippet cannot change the document's score. 

We propose an architecture for joint document and snippet ranking, i.e., stages (ii) and (iii), which leverages the intuition that relevant documents have good snippets and good snippets come from relevant documents. 
We note that modern web search engines display the most relevant snippets of the top-ranked documents to help users quickly identify truly relevant documents and answers \cite{sultan_et_al_Sentence_Ranking,Xu_et_al_Passage_Ranking,Yang_et_al_BERTserini}. The top-ranked snippets can also be used as a starting point for multi-document query-focused summarization, as in the \bioasq challenge \cite{Tsatsaronis_et_al_bioasq}. 
Hence, methods that identify good snippets are useful in several other applications, apart from \qa.
We also note that many neural models for stage (iv) have been proposed, often called \qa or Machine Reading Comprehension (\mrc) models \cite{Kadlec2016TextUW,Cui2017AttentionoverAttentionNN,Zhang2020RetrospectiveRF}, but they typically search for answers only in a particular, usually paragraph-sized snippet, which is given per question. For \qa systems that search large document collections, stages (ii) and (iii) are also important, if not more important, but have been studied much less in recent years, and not in a single joint neural model. 

The proposed joint architecture is general and can be used in conjunction with any neural text relevance ranker \cite{Mitra2018}. Given a query and $N$ possibly relevant documents from stage (i), the neural text relevance ranker scores all the snippets of the $N$ documents. Additional neural layers re-compute the score (ranking) of 
each document from the scores of its snippets. Other layers then revise the scores of the snippets taking into account the new scores of the documents. The entire model is trained to jointly predict document and snippet relevance scores. We experiment with two main instantiations of the proposed architecture, using \positdrmm \cite{McDonald_et_al_pdrmm}, hereafter called \pdrmm, as the neural text ranker, or a \bert-based ranker \cite{Devlin_et_al_bert}. 
We show how both \pdrmm and \bert can be used to score documents and snippets in pipelines, then how our architecture can turn them into models that jointly score documents and snippets.

Experimental results on biomedical data from \bioasq \cite{Tsatsaronis_et_al_bioasq} show the joint models vastly outperform the corresponding pipelines in snippet extraction, with fewer trainable parameters. Although our joint architecture is engineered to favor retrieving good snippets (as a near-final stage of \qa), results show that the joint models are also competitive in document retrieval. We also show that our joint version of \pdrmm, which has the fewest parameters of all models and does not use \bert, is competitive to \bert-based models, while also outperforming the best system of \bioasq 6 \cite{brokos_et_al} in both document and snippet retrieval. 
These claims are also supported by human evaluation on two test batches of \bioasq 7 (2019). 
To test our key findings on another dataset, we modified Natural Questions  \cite{Kwiatkowski_et_al_Natural_Questions}, which only includes questions and answer spans from a single document, so that it can be used for document and snippet retrieval. Again, our joint \pdrmm-based model largely outperforms the corresponding pipeline in snippet retrieval on the modified Natural Questions, though it does not perform better than the pipeline in document retrieval, since the joint model is geared towards snippet retrieval, i.e., even though it is forced to extract snippets from fewer relevant documents. 
Finally, we show that all the neural pipelines and joint models we considered improve the \bmtf ranking of traditional \IR on both datasets. 
We make our code and the modified Natural Questions publicly available.\footnote{See \url{http://nlp.cs.aueb.gr/publications.html} for links to the code and data.} 

\section{Methods} \label{sec:Methods}

\subsection{Document Ranking with PDRMM} \label{sec:pdrmm}

Our starting point is \positdrmm \cite{McDonald_et_al_pdrmm}, or \textbf{\pdrmm}, a differentiable extension of \drmm \cite{Guo_et_al_drmm} that obtained the best document retrieval results in \bioasq 6 \cite{brokos_et_al}. \newcite{McDonald_et_al_pdrmm} also reported it performed better than \drmm and several other neural rankers, including \pacrr \cite{Hui_et_al_pacrr}.

\begin{figure}[t]
\centering
\includegraphics[width=3in]{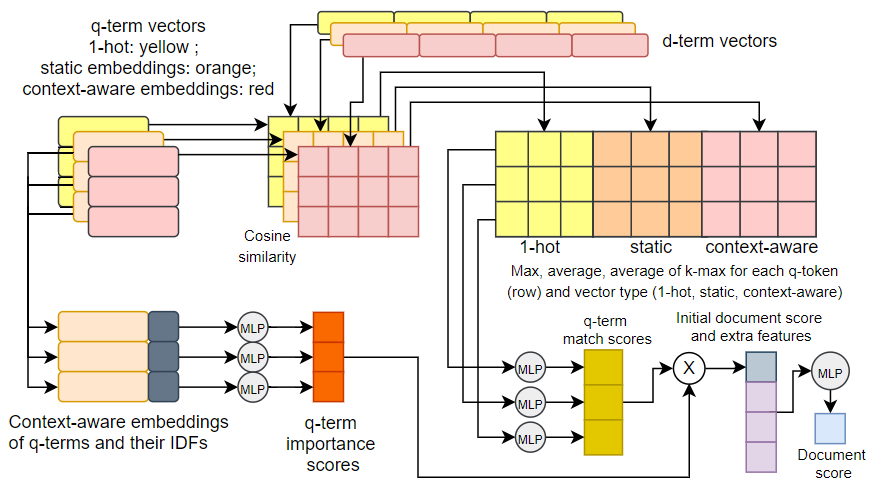}
\caption{\pdrmm for \emph{document} scoring. The same model (with different trained parameters) also scores \emph{sentences} in 
the  \pdrmmPipe pipeline and the joint \jpdrmm model (adding the layers of Fig.~\ref{fig:top-level-drmm}).
}
\label{fig:architecture}
\end{figure}

Given a query $q = \left<q_1, \dots, q_n\right>$ of $n$ query terms (\emph{q-terms}) and a document $d = \left<d_1, \dots, d_m\right>$ of $m$ terms (\emph{d-terms}), \pdrmm computes context-sensitive term embeddings $c(q_i)$ and $c(d_i)$ from the static (e.g., \wordtovec) embeddings $e(q_i)$ and $e(d_i)$ by applying two stacked convolutional layers with trigram filters, residuals \cite{He_et_al_residual}, and zero padding to $q$ and $d$, respectively.\footnote{\newcite{McDonald_et_al_pdrmm} use a \bilstm encoder instead of convolutions. We prefer the latter, because they are faster,
and we found that they do not degrade performance.}
\pdrmm then computes three similarity matrices $S_1, S_2, S_3$, each of dimensions $n \times m$ (Fig.~\ref{fig:architecture}). Each element $s_{i,j}$ of $S_1$ is the cosine similarity between $c(q_i)$ and $c(d_j)$. $S_2$ is similar, but uses the static word embeddings $e(q_i), e(d_j)$. $S_3$ uses one-hot vectors for $q_i, d_j$, signaling exact matches.
Three row-wise pooling operators are then applied to $S_1, S_2, S_3$: max-pooling (to obtain the similarity of the best match between the q-term of the row and any of the d-terms), average pooling (to obtain the average match), and average of $k$-max (to obtain the average similarity of the $k$ best matches).\footnote{We added average pooling to \pdrmm to balance the other pooling operators that favor long documents.} We thus obtain three scores from each row of each similarity matrix. By concatenating row-wise the scores from the three matrices, we obtain a new $n \times 9$ matrix $S'$   (Fig.~\ref{fig:architecture}). Each row of $S'$ indicates how well the corresponding q-term matched any of the d-terms, using the three different views of the terms (one-hot, static, context-aware embeddings). Each row of $S'$ is then passed to a Multi-Layer Perceptron (\mlp) to obtain a single match score per q-term.   

Each context aware q-term embedding is also concatenated with the corresponding \idf score (bottom left of Fig.~\ref{fig:architecture}) and passed to another \mlp that computes the importance of that q-term (words with low \idf{s} may be unimportant). 
Let $v$ be the vector containing the $n$ match scores of the q-terms, and $u$ the vector with the corresponding $n$ importance scores (bottom right of Fig.~\ref{fig:architecture}). The initial relevance score of the document is $\hat{r}(q, d) = v^T u$. Then $\hat{r}(q,d)$ is concatenated with four \emph{extra features}: z-score normalized \bmtf \cite{bm25}; percentage of q-terms with exact match in $d$ (regular and \idf weighted); percentage of q-term bigrams matched in $d$. An \mlp computes the final relevance $r(q,d)$ from the 5 features. 

Neural rankers typically re-rank the top $N$ documents of a conventional \IR system.
We use the same \bmtf-based \IR system as \newcite{McDonald_et_al_pdrmm}. \pdrmm is trained on triples $\left<q, d, d' \right>$, where $d$ is a relevant document from the top $N$ of $q$, and $d'$ is a random irrelevant document from the top $N$. We use hinge loss, requiring the relevance of $d$ to exceed that of $d'$ by a margin. 

\subsection{PDRMM-based Pipelines for Document and Snippet Ranking}
\label{sec:pdrmm_pipeline}

\newcite{brokos_et_al} used the `basic \cnn' (\bcnn) of \newcite{Yin_et_al_abcnn} to score (rank) the sentences of the re-ranked top $N$ documents. The resulting pipeline, \textbf{\pdrmmbcnn}, had the best document and snippet results in \bioasq 6, where snippets were sentences. Hence, \pdrmmbcnn is a reasonable  document and snippet retrieval baseline pipeline. In another pipeline, \textbf{\pdrmmPipe}, we replace \bcnn by a second instance of \pdrmm that scores sentences. The second \pdrmm instance is the same as when scoring documents (Fig.~\ref{fig:architecture}), but the input is now the query ($q$) and a single sentence ($s$). Given a triple $\left<q, d, d' \right>$ used to train the document-scoring \pdrmm, the sentence-scoring \pdrmm is trained to predict the true class (relevant,  irrelevant) of each sentence in $d$ and $d'$ using cross entropy loss (with a sigmoid on $r(q,s)$). As when scoring documents, the initial relevance score $\hat{r}(q,s)$ is combined with extra features using an \mlp, to obtain $r(q,s)$. The extra features are now different: character length of $q$ and $s$, number of shared tokens of $q$ and $s$ (with/without stop-words), sum of \idf scores of shared tokens (with/without stop-words), sum of \idf scores of shared tokens divided by sum of \idf scores of q-terms, number of shared token bigrams of $q$ and $s$, \bmtf score of $s$ against the sentences of $d$ and $d'$, \bmtf score of the document ($d$ or $d'$) that contained $s$. The two \pdrmm instances are trained separately.

\subsection{Joint PDRMM-based Models for Document and Snippet Ranking}
\label{sec:JPDRMM}

\begin{figure}[tb] 
\centering
\includegraphics[width=3in]{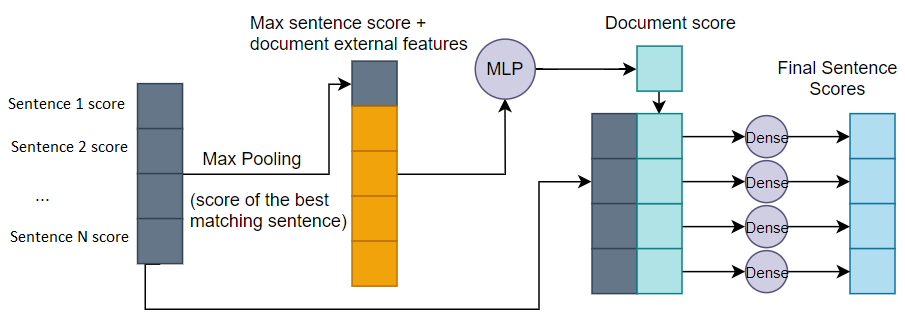}
\caption{Final layers of \jpdrmm and \jbert. The input sentence scores are generated by \pdrmm (Fig.~\ref{fig:architecture}) or \bert (Fig.~\ref{fig:BERT}) now applied to document \emph{sentences}. The document's score is obtained from the score of its best sentence and external features, and is also used to revise the sentence scores. 
Training jointly minimizes document and sentence loss.
}
\label{fig:top-level-drmm}
\end{figure}

Given a document $d$ with sentences $s_1, \dots, s_k$ and a query $q$, the joint document/snippet ranking version of \pdrmm, called \textbf{\jpdrmm}, processes separately each sentence $s_i$ of $d$, producing a relevance score $r(q, s_i)$ per sentence, as when \pdrmm scores sentences in the \pdrmmPipe pipeline. The highest sentence score $\max_i r(q, s_i)$ is concatenated (Fig.~\ref{fig:top-level-drmm}) with the extra features that are used when \pdrmm ranks documents, and an \mlp produces the document's score.\footnote{We also tried alternative mechanisms to obtain the document score from the sentence scores, including average of $k$-max sentence scores and hierarchical \rnn{s} \cite{yang_etal_2016_hierarchical}, but they led to no improvement.} \jpdrmm then revises the sentence scores, by concatenating the score of each sentence with the document score and passing each pair of scores to a dense layer to compute a linear combination, which becomes the revised sentence score. Notice that \jpdrmm is mostly based on scoring sentences, since the main goal for \qa is to obtain good snippets (almost final answers). The document score is obtained from the score of the document's best sentence (and external features), but the sentence scores are revised, once the document score has been obtained. We use sentence-sized snippets, for compatibility with \bioasq, but other snippet granularities (e.g., paragraph-sized) could also be used. 

\jpdrmm is trained on triples $\left< q, d, d'\right>$, where $d, d'$ are relevant and irrelevant documents, respectively, from the top $N$ of query $q$, as in the original \pdrmm; the ground truth now also indicates which sentences of the documents are relevant or irrelevant, as when training \pdrmm to score sentences in \pdrmmPipe. We sum the hinge loss of $d$ and $d'$ and the cross-entropy loss of each sentence.\footnote{Additional experiments with \jpdrmm, reported in the appendix, indicate that further performance gains are possible by tuning the weights of the two losses. 
} 
We also experiment with a \jpdrmm version that uses a pre-trained \bert model \cite{Devlin_et_al_bert} to obtain input token embeddings (of wordpieces) instead of the more conventional pre-trained (e.g., \wordtovec) word embeddings that \jpdrmm uses otherwise.
We call it \textbf{\bertjpdrmm} if \bert is fine-tuned when training \jpdrmm, and \textbf{\bertjpdrmmnf} if \bert is not fine-tuned. In another variant of \bertjpdrmm, called \textbf{\bertjpdrmmadapt}, the input embedding of each token is a linear combination of all the embeddings that \bert produces for that token at its different Transformer layers. The weights of the linear combination are learned via backpropagation.
This allows \bertjpdrmmadapt to learn which \bert layers it should mostly rely on when obtaining token embeddings. Previous work has reported that representations from different \bert layers may be more appropriate for different tasks \cite{Rogers2020}. \textbf{\bertjpdrmmadaptnf} is the same as \bertjpdrmmadapt, but \bert is not fine-tuned; the weights of the linear combination of embeddings from \bert layers are still learned. 

\subsection{Pipelines and Joint Models Based on Ranking with BERT}
\label{sec:BERT}

The \bertjpdrmm model we discussed above and its variants are essentially still \jpdrmm, which in turn invokes the \pdrmm ranker (Fig.~\ref{fig:architecture}, \ref{fig:top-level-drmm}); \bert is used only to obtain token embeddings that are fed to \jpdrmm. Instead, in this subsection we use  \bert as a ranker, replacing \pdrmm.

\begin{figure}[tb] 
\centering
\includegraphics[width=2.8in,height=2.8cm]{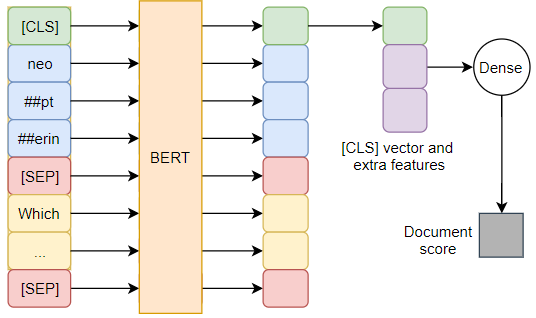}
\caption{\emph{Document} scoring with \bert. The same model scores \emph{sentences} in
\jbert (adding the layers of Fig.~\ref{fig:top-level-drmm}), but with an \mlp replacing the final dense layer.
}
\label{fig:BERT}
\end{figure}

For document ranking alone (when not cosidering snippets), we feed \bert with pairs of questions and documents (Fig.~\ref{fig:BERT}). \bert's top-layer embedding of the `classification' token \textsc{[cls]} is concatenated with external features (the same as when scoring documents with \pdrmm, Section~\ref{sec:pdrmm}), and a dense layer again produces the document's score. We fine-tune the entire model using triples $\left<q, d, d' \right>$ with a hinge loss between $d$ and $d'$, as when training \pdrmm to score documents.\footnote{We use the pre-trained uncased \bert \textsc{base} of \newcite{Devlin_et_al_bert}. The `documents' of the \bioasq dataset are concatenated titles and abstracts. Most question-document pairs do not exceed \bert's max.\ length limit of 512 wordpieces. If they do, we truncate documents. The same approach could be followed in the modified Natural Questions dataset,
where `documents' are Wikipedia paragraphs, 
but we did not experiment with \bert-based models on that dataset.
} 

Our two pipelines that use \bert for document ranking, \textbf{\bertbcnn} and \textbf{\bertpdrmm}, are the same as \pdrmmbcnn and \pdrmmPipe (Section~\ref{sec:pdrmm_pipeline}), respectively, but use the \bert ranker (Fig.~\ref{fig:BERT}) to score documents, instead of \pdrmm. The joint \textbf{\jbert} model is the same as \jpdrmm, but uses the \bert ranker (Fig.~\ref{fig:BERT}), now applied to sentences, instead of \pdrmm (Fig.~\ref{fig:architecture}), to obtain the initial sentence scores. The top layers of Fig.~\ref{fig:top-level-drmm} are then used, as in all joint models, to obtain the document score from the sentence scores and revise the sentence scores.
Similarly to \bertjpdrmm, we also experimented with variations of \jbert, which do not fine-tune the parameters of \bert (\textbf{\jbertnf}), use a linear combination (with trainable weights) of the \textsc{[cls]} embeddings from all the \bert layers (\textbf{\jbertadapt}), or both (\textbf{\jbertadaptnf}).

\subsection{BM25+BM25 Baseline Pipeline}

We include a \bmtfbmtf pipeline to measure the improvement of the proposed models on conventional \IR engines.
This pipeline uses the question as a query to the \IR engine and selects the $N_d$ documents with the highest \bmtf scores.\footnote{In each experiment, the same \IR engine and \bmtf hyper-parameters are used in all other methods.
All \bmtf hyper-parameters are tuned on development data.}
The $N_d$ documents are then split into sentences and \bmtf is re-computed, this time over all the sentences of the $N_d$ documents, to retrieve the $N_s$ best sentences.

\section{Experiments} \label{sec:Experiments}

\subsection{Data and Experimental Setup} \label{sec:Data}

\paragraph{BioASQ data and setup} 
Following \newcite{McDonald_et_al_pdrmm} and \newcite{brokos_et_al}, we experiment with data from \bioasq \cite{Tsatsaronis_et_al_bioasq}, which provides English biomedical questions, relevant documents from \textsc{medline/pubmed}\footnote{\url{https://www.ncbi.nlm.nih.gov/pubmed}}, and relevant snippets (sentences), prepared by biomedical experts. 
This is the only previous large-scale \IR dataset we know of that includes both gold documents and gold snippets.
We use the \bioasq 7 (2019) training dataset, which contains 2,747 questions, with 11 gold documents and 14 gold snippets per question on average. We evaluate on test batches 1--5 (500 questions in total) of \bioasq 7.\footnote{\bioasq 8 (2020) was ongoing during this work, hence we could not use its data for comparisons. See also the discussion of \bioasq results after expert inspection in Section~\ref{sec:experimentalResults}.}
We measure Mean Average Precision (\MAP) \cite{IRBook} for document and snippet retrieval, which are the official \bioasq evaluation measures.
The document collection contains approx.\ 18\textsc{m} articles (concatenated titles and abstracts only, discarding articles with no abstracts) from the \textsc{medline/pubmed} `baseline' 2018 dataset. In \pdrmm and \bcnn, we use the biomedical \wordtovec embeddings of 
\newcite{McDonald_et_al_pdrmm}. We use the \textsc{galago}\footnote{\url{www.lemurproject.org/galago.php}} \IR engine to obtain the top $N = 100$ documents per query. After re-ranking, we return $N_d = 10$ documents and $N_s = 10$ sentences, as required by \bioasq.
We train using Adam \cite{Kingma_et_al_Adam}. Hyper-parameters were tuned on held-out validation data.

\paragraph{Natural Questions data and setup}
Even though there was no other large-scale \IR dataset providing multiple gold documents and snippets per question, we needed to test our best models on a second dataset, other than \bioasq.
Therefore we modified the Natural Questions dataset \cite{Kwiatkowski_et_al_Natural_Questions} to a format closer to \bioasq's.
Each instance of Natural Questions consists of an \textsc{html} document of Wikipedia and a question.
The answer to the question can always be found in the document as if a perfect retrieval engine were used.
A short span of \textsc{html} source code is annotated by humans as a `short answer' to the question. A longer span of \textsc{html} source code that includes the short answer is also annotated, as a `long answer'. The long answer is most commonly a paragraph of the Wikipedia page. In the original dataset, more than 300,000 questions are provided along with their corresponding Wikipedia \textsc{html} documents, short answer and long answer spans.
We modified Natural Questions to fit the \bioasq setting.
From every Wikipedia \textsc{html} document in the original dataset, we extracted the paragraphs and indexed each paragraph separately to an ElasticSearch\footnote{\url{www.elastic.co/products/elasticsearch}} index, which was then used as our retrieval engine.
We discarded all the tables and figures of the \textsc{html} documents and any question that was answered by a paragraph containing a table.
For every question, we apply a query to our retrieval engine and retrieve the first $N=100$ paragraphs.
We treat each paragraph as a document, similarly to the \bioasq setting.
For each question, the gold (correct) documents are the paragraphs (at most two per question) that were included in the long answers of the original dataset. The gold snippets are the sentences (at most two per question) that overlap with the short answers of the original dataset. 
We discard questions for which the retrieval engine did not manage to retrieve any of the gold paragraphs in its top 100 paragraphs.
We ended up with 110,589 questions and 2,684,631 indexed paragraphs.
Due to lack of computational resources, we only use 4,000 questions for training, 400 questions for development, and 400 questions for testing, but we make the entire modified Natural Questions dataset publicly available.
Hyper-parameters were again tuned on held-out validation data. All other settings were as in the \bioasq experiments. 

\subsection{Experimental Results} 
\label{sec:experimentalResults}

\begin{table*}[ht]
  \centering
   \resizebox{1\textwidth}{!}{
       {\small 
        \begin{tabular}{|c|c||c|c||c|c|}
        \hline
        \textbf{Method} & \textbf{Params} & \textbf{Doc. MAP (\%)} & \textbf{Snip. MAP (\%)} & \textbf{Doc. Recall@10(\%)} & \textbf{Snip. Recall@10(\%)} \\
    \hline
    \bmtf+\bmtf & 4      & 6.86 & 4.29 & 48.65 & 4.93 \\
    \hline
    \pdrmmbcnn   & 21.83k & \textbf{7.47} & 5.67 & 52.97 & 12.43\\
    \pdrmmPipe   & 11.39k & \textbf{\underline{7.47}} & \underline{9.16} & \underline{52.97} & \underline{18.43} \\
    \jpdrmm      & \textbf{5.79k} & \underline{6.69} & \textbf{\underline{15.72}} & \textbf{\underline{53.68}} & \textbf{\underline{18.83}} \\
    \hline
    \bertbcnn  & 109.5M & \textbf{8.79} & 6.07 & \textbf{55.73} & 13.05\\
    \bertpdrmm & 109.5M & \textbf{\underline{8.79}} & \underline{9.63} & \textbf{\underline{55.73}} & \underline{19.30}\\
    \bertjpdrmm & 88.5M & \underline{7.59} & 16.82 & \underline{52.21} & 19.57 \\
    \bertjpdrmmadapt & 88.5M & 6.93 & 15.70 & 48.77 & 19.38 \\
    \bertjpdrmmnf & 3.5M & 6.84 & 15.77 & 48.81 & 17.95\\
    \bertjpdrmmadaptnf & 3.5M & 7.42 &  \textbf{\underline{17.35}} & 52.12 & \underline{19.66}\\
    \jbert     & 85M & \underline{7.93} & 16.29 & \underline{53.44} & \textbf{\underline{19.87}}\\
    \jbertadapt & 85M & 7.81 & 15.99 & 52.94 & \textbf{\underline{19.87}}\\
    \jbertnf   & \textbf{6.3K} & 7.90 & 15.99 & 52.78 & 19.64 \\
    \jbertadaptnf & \textbf{6.3K} & 7.84 & \underline{16.53} & 53.18 & 19.64 \\
    \hline
    Oracle & 0 & 19.24 & 25.18 & 72.67 & 41.14 \\
    Sentence \pdrmm & 5.68K & 6.39 & 8.73 & 48.60 & 18.57 \\
        \hline
        \end{tabular}
       }
   }
  \caption{Parameters learned, document and snippet \MAP on \textbf{\bioasq 7}, test batches 1--5, \textbf{before expert inspection}. 
  Systems in the 2nd (or 3rd) zone use (or not) \bert. In each zone, best scores shown in bold. In the 2nd and 3rd zones, we underline the results of the best pipeline, the results of \jpdrmm, and the best results of the \bertjpdrmm and \jbert variants. The differences between the underlined \MAP scores are statistically significant ($p \leq 0.01$). 
  }
  \label{tab:results_table}
\end{table*}

\paragraph{BioASQ results} 
Table~\ref{tab:results_table} reports document and snippet \MAP scores on the \bioasq dataset, along with the trainable parameters per method. For completeness, we also show recall at 10 scores, but we base the discussion below on \MAP, the official measure of \bioasq, which also considers the ranking of the 10 documents and snippets \bioasq allows participants to return.
The \textbf{Oracle} re-ranks the $N$ = 100 documents (or their snippets) that \bmtf retrieved, moving all the relevant documents (or snippets) to the top.
\textbf{Sentence \pdrmm} is an ablation of \jpdrmm without the top layers (Fig.~\ref{fig:top-level-drmm}); each sentence is scored using \pdrmm, then each document inherits the highest score of its snippets.

\pdrmmbcnn and \pdrmmPipe use the same document ranker, hence the document \MAP of these two pipelines is identical (7.47). However, \pdrmmPipe outperforms \pdrmmbcnn in snippet \MAP (9.16 to 5.67), even though \pdrmm has much fewer trainable parameters than \bcnn, confirming that \pdrmm can also score sentences and is a better sentence ranker than \bcnn. 
\pdrmmbcnn was the best system in \bioasq 6 for both documents and snippets, i.e., it is a strong baseline.
Replacing \pdrmm by \bert for document ranking in the two pipelines (\bertbcnn and \bertpdrmm) increases the document \MAP by 1.32 points (from 7.47 to 8.79) with a marginal increase in snippet \MAP for \bertpdrmm (9.16 to 9.63) and a slightly larger increase for \bertbcnn (5.67 to 6.07), at the expense of a massive increase in trainable parameters due to \bert (and computational cost to pre-train and fine-tune \bert).
We were unable to include a \bertPipe pipeline, which would use a second \bert ranker for sentences, with a total of approx.\ 220M trainable parameters, due to lack of computational resources. 

The main joint models (\jpdrmm, \bertjpdrmm, \jbert) vastly outperform the pipelines in snippet extraction, the main goal for \qa
(obtaining 
15.72, 16.82, 16.29 snippet \MAP, respectively), though their document \MAP is slightly lower (6.69, 7.59, 7.93) compared to the pipelines (7.47, 8.79), but still competitive. This is not surprising, since the joint models are geared towards snippet retrieval (they directly score sentences, document scores are obtained from sentence scores).
Human inspection of the retrieved documents and snippets, 
discussed below (Table~\ref{tab:bioasq_human_eval}), 
reveals that the document \MAP of \jpdrmm is actually higher than 
that of the best pipeline (\bertpdrmm), but is penalized in Table~\ref{tab:results_table} because of missing gold documents.

\jpdrmm, which has the fewest parameters of all neural models and does not use \bert at all, is competitive in snippet retrieval with models that employ \bert. 
More generally, the joint models use fewer parameters than comparable pipelines (see the zones of Table~\ref{tab:results_table}). Not fine-tuning \bert (\textsc{-nf} variants) leads to a further dramatic decrease in trainable parameters, at the expense of slightly lower document and snippet \MAP (7.59 to 6.84, and 16.82 to 15.77, respectively, for \bertjpdrmm, and similarly for \jbert). Using linear combinations of token embeddings from all \bert layers (\textsc{-adapt} variants) harms both document and snippet \MAP when fine-tuning \bert, but is beneficial in most cases when not fine-tuning \bert (\textsc{-nf}). The snippet \MAP of \bertjpdrmmnf increases from 15.77 to 17.35, and document \MAP increases from 6.84 to 7.42. A similar increase is observed in the snippet \MAP of \jbertnf (15.99 to 16.53), but \MAP decreases (7.90 to 7.84). In the second and third result zones of Table~\ref{tab:results_table}, we underline the results of the best pipelines, the results of \jpdrmm, and the results of the best \bertjpdrmm and \jbert variant. In each zone and column, the differences between the underlined \MAP scores are statistically significant ($p \leq 0.01$); we used single-tailed Approximate Randomization \cite{P18-1128}, 10k iterations, randomly swapping in each iteration the rankings of 50\% of queries. 
Removing the top layers of \jpdrmm (Sentence \pdrmm), clearly harms performance for both documents and snippets.
The oracle scores indicate  there is still scope for improvements in both documents and snippets.

\paragraph{BioASQ results after expert inspection}

\begin{table*}[ht]
  \centering
  \resizebox{0.9\textwidth}{!}{
    {\small
      \begin{tabular}{|c||c|c||c|c|}
    \hline
     &
    \multicolumn{2}{|c||}{\textbf{Before expert inspection}} &
    \multicolumn{2}{|c|}{\textbf{After expert inspection}} 
    \\
    \hline
    \textbf{Method} & \textbf{Document \MAP} & \textbf{Snippet \MAP}
    & \textbf{Document \MAP} & \textbf{Snippet \MAP} 
    \\
    \hline
    \bertpdrmm      & \textbf{7.29} & 7.58  
    & 14.86 & 15.61 
    \\
    \jpdrmm         & 5.16 & 12.45 
    & \textbf{16.55} & 21.98 
    \\
    \bertjpdrmmnf & 6.18 & \textbf{13.89} 
    & 14.65 & \textbf{23.96} 
    \\
    \hline
    Best \bioasq 7 competitor & n/a & n/a & 13.18 & 14.98 \\
    \hline
    \end{tabular}
    }
  }
  \caption{Document and snippet \MAP (\%)  on \textbf{\bioasq 7 test batches 4 and 5} \textbf{before and after post-contest expert inspection} of system responses, for methods that participated in \bioasq 7. We also show the results (after inspection) of the \textbf{best other participants of \bioasq 7} for the same batches.}
  \label{tab:bioasq_human_eval}
\end{table*}

\begin{table*}[ht]
  \centering
  \resizebox{0.8\textwidth}{!}{
      {\small
      \begin{tabular}{|c|c|c|c|c|c|c|}
        \hline
        &
        \multicolumn{3}{|c|}{\textbf{Document Retrieval}} &
        \multicolumn{3}{|c|}{\textbf{Snippet Retrieval}} \\
        \hline
        \textbf{Method} & 
        \textbf{MRR} & \textbf{Recall@1} & \textbf{Recall@2} & \textbf{MRR}  & \textbf{Recall@1} & \textbf{Recall@2}\\
        \hline
        \bmtfbmtf & 30.18 & 16.50 & 29.75 & 8.19 & 3.75 & 7.13\\
        \pdrmmPipe & \textbf{40.33} & \textbf{28.25} & \textbf{38.50} & 22.86 & 13.75 & 22.75\\
        \jpdrmm & 36.50 & 24.50 & 36.00 & \textbf{26.92} & \textbf{19.00} & \textbf{25.25}\\
        \hline
        \end{tabular}
        }
  }
  \caption{\MRR (\%) and recall at top 1 and 2 (\%) on the modified \textbf{Natural Questions} dataset. 
  }
  \label{tab:NQ_results}
\end{table*}

At the end of each \bioasq annual contest, the biomedical experts who prepared the questions and their gold documents and snippets inspect the  responses of the participants. If any of the documents and snippets returned by the participants are judged relevant to the corresponding questions, they are added to the gold responses. This process enhances the gold responses and avoids penalizing participants for responses that are actually relevant, but had been missed by the experts in the initial gold responses. However, it is unfair to use the post-contest enhanced gold responses to compare systems that participated in the contest to systems that did not, because the latter may also return documents and snippets that are actually relevant and are not included in the gold data, but the experts do not see these responses and they are not included in the gold ones. The results of Table~\ref{tab:results_table} were computed on the initial gold responses of \bioasq 7, before the post-contest revision, because not all of the methods of that table participated in \bioasq 7.\footnote{Results \emph{without} expert inspection can be obtained at any time, using the \bioasq evaluation platform. Results with expert inspection can only be obtained during the challenge.} In Table~\ref{tab:bioasq_human_eval}, we show results on the revised post-contest gold responses of \bioasq 7, for those of our methods that participated in the challenge. We show results on test batches 4 and 5 only (out of 5 batches in total), because these were the only two batches were all three of our methods participated together. Each batch comprises 100 questions. We also show the best results (after inspection) of our competitors in \bioasq 7, for the same batches.

A first striking observation in Table~\ref{tab:bioasq_human_eval} is that 
all results improve substantially after expert inspection, i.e., all systems retrieved many relevant documents and snippets the experts had missed.
Again, the two joint models (\jpdrmm, \bertjpdrmmnf) vastly outperform the \bertpdrmm pipeline in snippet \MAP. As in Table~\ref{tab:results_table}, before expert inspection the pipeline has slightly better document \MAP than the joint models. However, 
after expert inspection \jpdrmm exceeds the pipeline in document \MAP by almost two points.
\bertjpdrmmnf performs two points better than \jpdrmm in snippet \MAP after expert inspection, though \jpdrmm performs two points better in document \MAP. After inspection, the document \MAP of \bertjpdrmmnf is also very close to the pipeline's. Table~\ref{tab:bioasq_human_eval} confirms that \jpdrmm is competitive with models that use \bert, despite having the fewest parameters. 
All of our methods clearly outperformed the competition.

\paragraph{Natural Questions results}
Table~\ref{tab:NQ_results} reports results on the modified Natural Questions dataset.
We experiment with the best pipeline and joint model of  Table~\ref{tab:results_table} that did not use \bert (and are computationally much cheaper), i.e., \pdrmmPipe and \jpdrmm, comparing them to the more conventional \bmtfbmtf baseline.
Since there are at most two relevant documents and snippets per question in this dataset, we measure Mean Reciprocal Rank (\MRR) \cite{IRBook}, and Recall at top 1 and 2.
Both \pdrmmPipe and \jpdrmm clearly outperform the \bmtfbmtf pipeline in both document and snippet retrieval.
As in Table~\ref{tab:results_table}, the joint \jpdrmm model outperforms the \pdrmmPipe pipeline in snippet retrieval, but the pipeline performs better in document retrieval. Again, this is unsurprising, since the joint models are geared towards snippet retrieval. 
We also note that \jpdrmm uses half of the trainable parameters of \pdrmmPipe
(Table~\ref{tab:results_table}). No comparison to previous work that used the original Natural Questions is possible, since the original dataset provides a single document per query (Section~\ref{sec:Data}).

\vspace*{-1mm}
\section{Related Work}
\label{sec:relatedWork}
\vspace*{-1mm}

Neural document ranking \cite{Guo_et_al_drmm,Hui_et_al_pacrr,Pang_et_al_reviewer_6,copacrr,McDonald_et_al_pdrmm} only recently managed to improve the rankings of conventional \IR; see \newcite{NeuralHype} for caveats. Document or passage ranking models based on \bert have also been proposed, with promising results, but most use only simplistic task-specific layers on top of \bert \cite{Yang_et_al_BERT_for_Ad_Hoc_Document_Retrieval,Nogueira_et_al_Passage_Re_ranking_with_BERT}, similar to our use of \bert for document scoring (Fig.~\ref{fig:BERT}). An exception is the work of \newcite{MacAvaney_et_al_trec_doc_ranking}, who explored combining \elmo \cite{elmo} and \bert \cite{Devlin_et_al_bert} with complex neural \IR models, namely \pacrr \cite{Hui_et_al_pacrr}, \drmm \cite{Guo_et_al_drmm}, \knrm \cite{Dai_et_al_knrm}, \convknrm \cite{convknrm}, an approach that we also explored here by combining \bert with \pdrmm in \bertjpdrmm and \jbert. However, we retrieve both documents and snippets, whereas \newcite{MacAvaney_et_al_trec_doc_ranking} retrieve only documents. 

Models that directly retrieve documents by indexing neural document representations, rather than re-ranking documents retrieved by conventional \IR, have also been proposed \cite{Fan_et_al_reviewer_3,Ai_et_al_reviewer_5,khattab_et_al_2020_colbert}, but none addresses both document and snippet retrieval. \newcite{Yang_et_al_BERTserini} use \bert to encode, index, and directly retrieve snippets, but do not consider documents; indexing snippets is also computationally costly.
\newcite{lee-etal-2019-latent} propose a joint model for direct snippet retrieval (and indexing) and answer span selection, again without retrieving documents.

No previous work combined document and snippet retrieval in a joint neural model.
This may be due to existing datasets, which do not provide both gold documents and gold snippets, with the exception of \bioasq, which is however small by today's standards (2.7k training questions, Section~\ref{sec:Data}). 
For example, 
\newcite{Pang_et_al_reviewer_6} used much larger clickthrough datasets from a Chinese search engine, as well as datasets from the 2007 and 2008 \textsc{trec} Million Query tracks \cite{Tao_et_al_2010_LETOR}, but these datasets do not contain gold snippets. 
\squad \cite{squad} and \squad v.2 \cite{squad2} provide 100k and 150k questions, respectively, but for each question they require extracting an exact answer span from a single given Wikipedia paragraph; no snippet retrieval is performed, because the relevant (paragraph-sized) snippet is given. \newcite{ahmad-etal-2019-reqa} provide modified versions of \squad and Natural Questions, suitable for direct snippet retrieval, but do not consider document retrieval. 
SearchQA \cite{Dunn_et_al_SearchQA} provides 140k questions, along with 50 snippets per question.
The web pages the snippets were extracted from, however, are not included in the dataset, only their \textsc{url}s, and crawling them may produce different document collections, since the contents of web pages often change, pages are removed etc. 
\textsc{ms-marco} \cite{Nguyen_et_al_MS_MARCO} was constructed using 1M queries  extracted from Bing's logs. For each question, the dataset includes the snippets returned by the search engine for the top-10 ranked web pages. 
However the gold answers to the questions 
are not spans of particular retrieved snippets, but were freely written by humans after reading the returned snippets.
Hence, gold relevant snippets (or sentences) cannot be identified, 
making this dataset unsuitable for our purposes.

\section{Conclusions and Future Work}

Our contributions can be summarized as follows: (1) We proposed an architecture to jointly rank documents and snippets with respect to a question, two particularly important stages in \qa for large document collections; our architecture can be used with any neural text relevance model. (2) We instantiated the proposed architecture using a recent neural relevance model (\pdrmm) and a \bert-based ranker. (3) Using biomedical data (from \bioasq), we showed that the two resulting joint models (\pdrmm-based and \bert-based) vastly outperform the corresponding pipelines in snippet retrieval, the main goal in \qa for document collections, using fewer parameters, and also remaining competitive in document retrieval. (4) We showed that the joint model (\pdrmm-based) that does not use \bert is competitive with \bert-based models, outperforming the best \bioasq 6 system; our joint models (\pdrmm- and \bert-based) also outperformed all \bioasq 7 competitors. (5) We provide a modified version of the Natural Questions dataset, suitable for document and snippet retrieval. (6) We showed that our joint \pdrmm-based model also largely outperforms the corresponding pipeline on open-domain data (Natural Questions) in snippet retrieval, even though it performs worse than the pipeline in document retrieval. (7) We showed that all the neural pipelines and joint models we considered improve the traditional \bmtf ranking on both datasets. (8) We make our code publicly available.

We hope to extend our models and datasets for stage (iv), i.e., to also identify exact answer spans within snippets (paragraphs), similar to the answer spans of \squad \cite{squad,squad2}.
This would lead to a multi-granular retrieval task, where systems would have to retrieve relevant documents, relevant snippets, and exact answer spans from the relevant snippets. 
\bioasq already includes this multi-granular task, but exact answers are provided only for factoid questions and they are freely written by humans, as in \textsc{ms-marco}, with similar limitations.
Hence, appropriately modified versions of the \bioasq datasets are needed.

\section*{Acknowledgements}
We thank Ryan McDonald for his advice in earlier stages of this work. 

\bibliographystyle{acl_natbib}
\bibliography{acl2021}

\section*{Appendix}
\label{sec:Appendix}

\subsection*{Tuning the weights of the two losses and the effect of extra features in \jpdrmm}

In Table~\ref{tab:results_table}, all  joint models used the sum of the document and snippet loss ($L = L_{\textit{doc}} + L_{\textit{snip}}$). By contrast, in Table~\ref{tab:crossvalid_10fold_jpdrmm} we use a linear combination $L = L_{\textit{doc}} + \lambda_{\textit{snip}} L_{\textit{snip}}$ and tune the hyper-parameter $\lambda_{\textit{snip}} \in \{10, 1, 0.1, 0.01\}$. We also try removing the extra document and/or sentence features (Fig.~\ref{fig:architecture}--\ref{fig:BERT}) to check their effect. 
This experiment was performed only with \jpdrmm, which is one of our best joint models and computationally much cheaper than methods that employ \bert. As in Table~\ref{tab:results_table}, we use the \bioasq data, but here we perform a 10-fold cross-validation on the union of the training and development subsets. This is why the results for $ \lambda_{\textit{snip}} = 1$ when using both the sentence and document extra features (row 4, in italics) are slightly different than the corresponding \jpdrmm results of Table~\ref{tab:results_table} (6.69 and 15.72, respectively). 

\begin{table}[ht]
  \centering
   \resizebox{\columnwidth}{!}{
       {\small 
        \begin{tabular}{|c|c|c|c|c|}
        \hline
        \textbf{Sent.} & \textbf{Doc.} &&  \textbf{Doc.} & \textbf{Snip.} \\
        \textbf{Extra} & 
        \textbf{Extra} & 
        \textbf{$\lambda_{\textit{snip}}$} & 
        \textbf{MAP (\%)} & 
        \textbf{MAP (\%)}
        \\
    \hline
    Yes & Yes & 10 & 6.23 $\pm$ 0.14 & 14.73 $\pm$ 0.32 \\
    Yes & No  & 10 & 1.20 $\pm$ 0.14 & 3.59  $\pm$ 0.45 \\
    No  & Yes & 10 & 1.18 $\pm$ 0.23 & 2.19  $\pm$ 0.29 \\
    \textit{Yes} & \textit{Yes} & \textit{1}  & \textit{6.80 $\pm$ 0.07} & \textit{15.42 $\pm$ 0.23} \\
    Yes & No  & 1  & 1.35 $\pm$ 0.24 & 3.77 $\pm$ 0.73 \\
    No  & Yes & 1  & 7.35 $\pm$ 0.16 & 14.58 $\pm$ 0.88 \\
    Yes & Yes & 0.1  & \textbf{7.85 $\pm$ 0.08} & 17.28 $\pm$ 0.26 \\
    Yes & No  & 0.1  & 6.77 $\pm$ 0.25 & 13.86 $\pm$ 1.10 \\
    No  & Yes & 0.1  & 7.59 $\pm$ 0.12 & 15.77 $\pm$ 0.60 \\
    Yes & Yes & 0.01 & 7.83 $\pm$ 0.07 & \textbf{17.34 $\pm$ 0.37} \\
    Yes & No  & 0.01 & 6.61 $\pm$ 0.19 & 12.96 $\pm$ 0.29 \\
    No  & Yes & 0.01 & 7.65 $\pm$ 0.10 & 14.24 $\pm$ 1.63 \\
    \hline
    \end{tabular}
       }
   }
  \caption{\textbf{\jpdrmm results on \bioasq 7} data for \textbf{tuned weights of the two losses}, \textbf{with and without the extra sentence and document features}. The 4th row (in italics) corresponds to the \jpdrmm configuration of Table~\ref{tab:results_table}, but the results here are slightly different, because we used a 10-fold cross-validation on the training and development data. The \MAP scores are averaged over the 10 folds. We also report standard deviations ($\pm$). 
  }
  \label{tab:crossvalid_10fold_jpdrmm}
\end{table}

Table~\ref{tab:crossvalid_10fold_jpdrmm} shows that further performance gains (6.80 to 7.85 document \MAP, 15.42 to 17.34 snippet \MAP) are possible by tuning the weights of the two losses.
The best scores are obtained when using both the extra sentence and document features.
However, the model performs reasonably well even when one of the two types of extra features is removed, with the exception of $ \lambda_{\textit{snip}} = 10$.
The standard deviations of the \MAP scores over the folds of the cross-validation indicate that the performance of the model is reasonably stable.

\subsection*{Error Analysis and Limitations}

We conducted an exploratory analysis of the retrieved snippets in the two datasets. For each dataset, we used the model with the best snippet retrieval performance, i.e., \jpdrmm for the modified Natural Questions (Table~\ref{tab:NQ_results}) and \bertjpdrmmadaptnf for \bioasq (Table~\ref{tab:results_table}).

Both models struggle to retrieve the gold sentences when the answer is not explicitly mentioned in them. 
For example, the gold sentence for the question \textit{``What is the most famous fountain in Rome?''} of the Natural Questions dataset is:

\smallskip
\noindent\textit{``The Trevi Fountain (Italian: Fontana di Trevi) is a fountain in the Trevi district in Rome, Italy, designed by Italian architect Nicola Salvi and completed by Giuseppe Pannini.''}
\smallskip

\noindent Instead, the top sentence of \jpdrmm is the following, which looks reasonably good, but mentions famous fountains (of a particular kind) \emph{near} Rome. 

\smallskip
\noindent\textit{``The most famous fountains of this kind were found in the Villa d'Este, at Tivoli near Rome, which featured a hillside of basins, fountains and jets of water, as well as a fountain which produced music by pouring water into a chamber, forcing air into a series of flute-like pipes.''}.
\smallskip

\noindent To prefer the gold sentence, the model needs to know that Fontana di Trevi is also very famous, but this information is not included in the gold sentence itself, though it is included in the next sentence:

\smallskip
\noindent\textit{``Standing 26.3 metres (86 ft) high and 49.15 metres (161.3 ft) wide, it is the largest Baroque fountain in the city and one of the most famous fountains in the world.''}
\smallskip

\noindent Hence, some form of multi-hop \qa \cite{yang-etal-2018-hotpotqa,bauer-etal-2018-commonsense,khot-etal-2019-whats,saxena-etal-2020-improving} seems to be needed to combine the information that Fontana di Trevi is in Rome (explicitly mentioned in the gold sentence) with information from the next sentence and, more generally, other sentences even from different documents. 

In the case of the question \textit{``What part of the body is affected by mesotheliomia?''} of the \bioasq dataset, the gold sentence is:

\smallskip
\noindent\textit{`'Malignant pleural mesothelioma (MPM) is a hard to treat malignancy arising from the mesothelial surface of the pleura.''
}
\smallskip

\noindent Instead, the top sentence of \bertjpdrmmadaptnf is the following, which contains several words of the question, but not `mesothelioma', which is the most important question term.

\smallskip
\noindent\textit{``For PTs specialized in acute care, geriatrics and pediatrics, the body part most commonly affected was the low back, while for PTs specialized in orthopedics and neurology, the body part most commonly affected was the neck.''}
\smallskip

\noindent In this case, the gold sentence does not explicitly convey that the pleura is a membrane that envelops each lung of the human body and, therefore, a part of the body. Again, this additional information can be found in other sentences.

\end{document}